# Through Silicon Via Aware Design Planning for Thermally Efficient 3-D Integrated Circuits

Yibo Chen, Eren Kursun, *Senior Member, IEEE,* Dave Motschman, Charles Johnson, and Yuan Xie, *Senior Member, IEEE*

*Abstract*—3-D integrated circuits (3-D ICs) offer performance advantages due to their increased bandwidth and reduced wirelength enabled by through-silicon-via structures (TSVs). Traditionally TSVs have been considered to improve the thermal conductivity in the vertical direction. However, the lateral thermal blockage effect becomes increasingly important for TSV via farms (a cluster of TSV vias used for signal bus connections between layers) because the TSV size and pitch continue to scale in $\mu$m range and the metal to insulator ratio becomes smaller. Consequently, dense TSV farms can create lateral thermal blockages in thinned silicon substrate and exacerbate the local hotspots. In this paper, we propose a thermal-aware via farm placement technique for 3-D ICs to minimize lateral heat blockages caused by dense signal bus TSV structures. By incorporating thermal conductivity profile of via farm blocks in the design flow and enabling placement/aspect ratio optimization, the corresponding hotspots can be minimized within the wirelength and area constraints.

*Index Terms*—3-D floorplanning, analysis, modeling, physical design

## I. INTRODUCTION

3D INTEGRATION offers a number of benefits for future VLSI design [1]–[7], such as 1) higher packing density and smaller footprint; 2) shorter global interconnect due to the short length of through-silicon vias (TSVs) and the flexibility of vertical routing, leading to higher performance and lower power consumption of interconnects [8]; 3) support of heterogeneous integration: each single die can have different technologies [9]. Increased interconnectivity is considered a key advantage in improving performance by leveraging the high-bandwidth and low-latency TSV structures [10]. TSVs are essential in implementing interconnectivity across layers in a wide range of 3-D partitioning approaches: from fine-grain partitioning options like SOC IP blocks [11], [12] to coarser/functional partitioning of processor-memory layers [13], [14].

TSV-based 3-D buses are usually clustered together as dense TSV via farm across the range of partitioning options [15], [16]. As buses are widely used for intensive communications in multicore chips with high power density [17], [18]; recent studies have shown that 3-D integration exacerbates the on-chip temperatures due to the higher transistor density per cooling area and the increased vertical resistances. TSVs have been considered to help drain the heat in the vertical direction [19], [20]. As a result, thermal vias have been employed during placement and routing of 3-D ICs to help the vertical heat conduction [21]–[24].

However, as the TSV scaling continues in advanced 3-D technologies, the metal-to-insulator ratio and the distance between TSVs become smaller. While the vertical conductivity improvement is true for larger thermal vias, small and dense signal vias can act as thermal blockages to the neighboring blocks in thin silicon. The lateral thermal-blockage effects of TSV clusters (via-farms) become more prominent and can have a significant impact on the chip thermal profile. (Fig. 1(a) shows an illustrative case where a heavily interconnected SOC IP block is surrounded with TSV-farm structures).

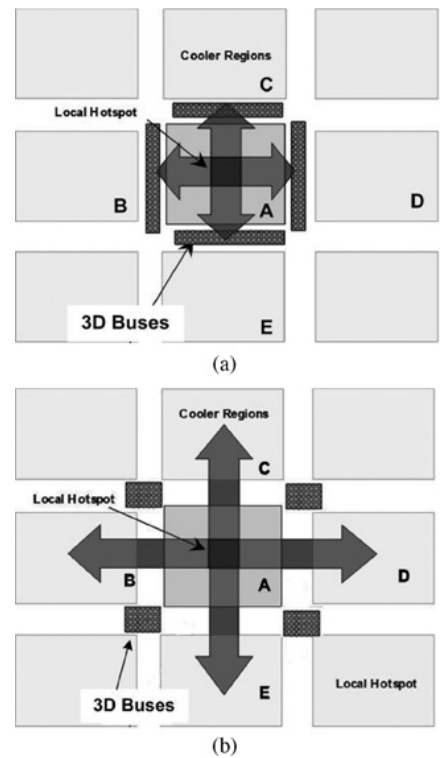

Fig. 1. Thermally optimized 3-D TSV-bus (or TSV via farm) structure with enhanced heat flow. (a) Original lateral heat flow where heat dissipation is blocked by the TSV-farm structures. (b) Improved lateral heat flow where the blockage effected is mitigated by the thermal-aware optimization. (a) Original lateral heat flow. (b) After optimized bus flow.

For most 3-D partitioning approaches 3-D floorplanning is usually the first step in the physical design flow [2], [25]. In the current 3-D design, flow thermal blockage implications of TSV structures are not taken into consideration. The majority of the existing studies on thermal-aware 3-D floorplanning methodologies [26]–[29] have mainly focused on the thermal effects of stacking functional units or IP cores, and some consider enhancing the vertical thermal conductivity with the help of thermal vias.

TSVs may have different effects on different device layers in the stack and their conductivities are highly dependent on the technology and design parameters. Current design tools are oblivious to the thermal conductivity range and differences of density via farms in x, y, z directions. For instance a 3-D bus can improve the vertical heat flow (in z direction), at the same time it can also create thermal blockage effects in the lateral (x, y) direction in the layers that it passes through. This effect can become prominent in dense via farms in thinned silicon. Consequently, in this paper, we study such lateral thermal-blockage effect for TSVs, and make the following contributions.

1) We observe that dense TSV regions alter the thermal profile of 3-D stacks. However, current design automation flows don't take this effect into consideration.
2) We show that TSV structures can affect the thermal profiles of the layers that they purely pass through (changing the thermal profiles of layers to which they are not even electrically connected). This requires specialized design planning approaches.
3) We then propose a thermal-blockage-aware design flow that incorporates a detailed thermal conductivity profile for individual via-farm structures to maximize the heat flow in the lateral and vertical directions (Section IV). The technique maximizes the lateral heat flow by removing the via-blockages on the heat dissipation paths. TSV-farms are treated as soft macros for temperature flow optimization. The aspect ratio and placement of the via-farm structures can be changed through design iterations to improve the thermal profile without causing any performance/area or wirelength degradation. The optimization flow targets maximum performance and minimum wirelength options, while improving the thermal profile at the same time.
4) The proposed technique can improve the performance of temperature-aware floorplanning techniques by minimizing the exacerbated heating caused by vertical interconnect. As TSVs affect the design flow in multiple device layers, a combination of both interlayer and intralayer optimization are needed to effectively deal with such cases. Our experimental results show that the peak/average temperature can be reduced significantly with such thermal-blockage-aware optimization. To the best of our knowledge, this is the first work to study and mitigate the thermal-blockage effects of the 3-D TSVs.

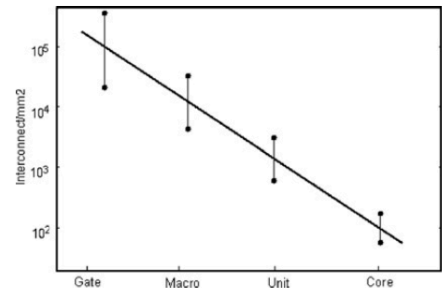

Fig. 2. Interconnect density for different partitioning approaches across 3-D spectrum [30].

## II. PRELIMINARIES ON INTERCONNECTS IN 3-D ICS

As discussed in the previous section 3-D integration provides a number of advantages, such as improved packaging density, reduced interconnect-length across the stack and heterogeneous integration of disparate technologies. The performance advantages associated with 3-D stacking have been illustrated by various studies [2]–[7], [12], [30].

Over the years TSV sizes have scaled to smaller diameter and pitch, providing higher density interconnects. This led to the feasibility of fine-grain partitioning options, such as core, unit-level approaches. The corresponding TSV size and interconnect density demands for various partitioning options differ significantly as Fig. 2 illustrates.

Even for coarser partitioning options, 3-D interconnect density (in interconnect per $mm^2$) can be sufficient to alter the thermal profile of the stack. Due to the natural clustering of signal TSVs corresponding thermal effects can alter the thermal profile of the device layers not only in the vertical direction but also in the lateral direction. While thermal via planning is extensively studied by various researchers, thermal effects of high density lateral thermal effects of TSV farms have not been explored from a design planning point of view.

## III. HEAT CONDUCTIVITY CHANGES DUE TO TSV FARMS

While providing additional advantages over 2-D counterparts, 3-D buses require further analysis and optimization as a result of the more complex interaction among performance, power, and temperature characteristics in 3-D. Traditionally, such TSV buses have been considered to help improve the thermal conductivity in the vertical direction, with the general perception that Copper/Tungsten TSVs have good thermal conductivities [19], [21], [23]. However, the lateral thermal blockage effects for TSV via farms have not been well-studied in the past. A number of factors contributing to the TSV lateral heat blockage effects are:

1) *TSV Technology:* Large copper TSVs (with diameters in tens of $\mu$m) have been shown to improve the vertical heat dissipation as in thermal vias [31]; however, dense and insulated tungsten via farms with few $\mu$m diameters exhibit different thermal characteristics according to our preliminary observation. They can even act as thermal blockages in the lateral direction in thinned device layers. In advanced 3-D technologies, the sizes and pitches of such TSVs can be as small as a few $\mu$m. While this density can be leveraged toward performance, the resulting thermal behavior deteriorates with smaller metal/insulator ratio and TSV-to-TSV distances. In particular, the electrical insulation in the TSVs has thermal insulation effect in the lateral direction due to the nature of the materials used for insulation. Such lateral thermal blockage effect can be even more serious in the unitlevel and core-level 3-D floorplanning, where the signal buses among units or cores are implemented as dense TSV via cluster (or TSV via farm).
2) *3-D Wafer Thinning:* In a 2-D setting, silicon layer thickness is frequently around $700 - 800$ $\mu$m, which enables effective heat dissipation in lateral and vertical direction in the silicon [31], [32]. The existing temperature gradient between Block A and Blocks B/D in Fig. 1(a) can be easily reduced because of the lateral heat dissipation in the $x/y$ directions in a 2-D architecture. However, in 3-D IC design, aggressive wafer thinning is usually needed to reduce the TSV via size and pitch while maintaining the TSV aspect ratio. For example, in 3-D stacked IC, the thickness of a wafer can be in the range of $10 - 100$ $\mu$m [2]. As a result, the lateral heat conductivity to the neighboring blocks is greatly diminished. The lateral heat gradients as well as hotspots are exacerbated. Such thermal implications may not be of serious concern for designs with uniform power density and relative low power dissipation, such as DRAM layers, however, the higher power processor or accelerator architectures are affected by the localized heating in the thinned processor layers. Hence, we focus our attention to the microprocessor stacks with local power density variations within the layers.

Fig. 3 shows an illustrative case of a dense TSV farm, where the TSV-farm have different characteristics. Compared to Fig. 3(b), the via-farm area in Fig. 3(c) creates a blockage to the heat flow due to the lower conductivity of the via-structures in addition to the electrical insulation, with six degrees of temperature increases. The local hotspot exacerbates, also the span of the hotspot area has expanded as shown in the corresponding thermal map.

Such lateral thermal-blockage effects of TSV via farms can have significant impact for the coarse-granularity 3-D SoC design or 3-D multicore design, with either unit-level or corelevel partitioning.

1) *Unit-Level Case:* Recent studies have shown heavily interconnected structures, such as instruction scheduling units, register files, and integer/floating point execution units as frequent hotspots within the core [17]. Such structures have heavy communication demands to the rest of the core, as well as high power densities. In a 3-D setting TSV farms surrounding such units can cause increased hotspot temperatures by creating lateral heat blockages.

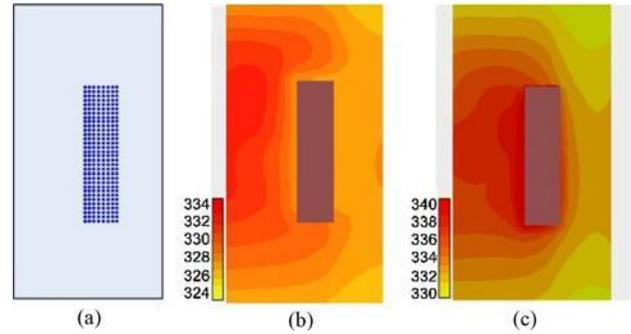

Fig. 3. Thermal maps of alternative TSV-farm structures in thinned 10$\mu$m layer. (a) Top view of the dense via farm. (b) Highly conductive copper TSV implementation. (c) Thermal blockage due to the composite, insulated Tungsten via farm, with six degrees of temperature increases.

2) *Core-Level Case:* At the core-level, temperature differences have been reported to be high where highactivity cores communicate with the other layers through the surrounding TSV farm/bus structures. Since the proximity and placement of bus structures will cause unexpected increases in the hotspot temperatures, even identical workloads can result in different temperature profiles depending on the bus-blockages. This impacts the effectiveness of workload-aware management algorithms as well since such heating is not predictable at system-level.

Thermal behavior of 3-D via-farm structures varies significantly with the TSV size/pitch and material selection in the corresponding 3-D manufacturing flow, as well as the specific geometry and placement decisions at the design time. The localized hotspots due to such lateral heat blockage are difficult to predict and can have detrimental effects on not only energy efficiency and thermal profile of the 3-D stack but also reliability characteristics. When wide 3-D bus structures with dense vias span multiple-layers vertically they can create thermal blockages throughout the 3-D stack. Such blockages can affect multiple layers in the stack even those that are not directly connected to the via farms, which consists of 3-D bus signals.

Consequently, with thermal characterization and modeling for TSV buses (via farm) (see Section III), we propose thermal-blockage-aware bus optimization techniques, such that the temperature profile of the individual layers as well as the overall stack can be improved. For example, the floorplan of Fig. 1(a) can be optimized to help lateral thermal dissipation as shown in Fig. 1(b). By identifying the maximum temperature gradients in the layer, the bus structures are removed from these efficient heat dissipation paths. The aspect ratios are adjusted for minimizing the heat blockages in the correspond-ing directions targeting localized hotspots. As a result, the temperature profile and energy efficiency are improved without compromising the performance advantages of vertical bus structures.

TABLE I
Thermal Conductivities of Materials in 3-D ICs

| Material | Thermal Conductivity (W/(mK)) |
|---|---|
| Copper (TSV) | 401 |
| Tungsten (TSV) | 173 |
| Silicon | 149 |
| Polysilicon | 23.1 |
| Thermal interface material [35] | 5.0 |
| SiO$_2$ (Dielectric / TSV liner [36], [37]) | 1.38 |
| Bonding adhesive [38] | 0.29 |

## IV. THERMAL CHARACTERIZATION AND MODELING OF 3-D TSV-FARM STRUCTURES

In this section, we present the characterization and modeling of TSV-Farm structure, which are used to guide our optimization techniques that are presented in the next section.

### A. Thermal Analysis Methodology

We use a combination of commercial multiphysics solvers for the stack thermal model including the TSV-farm structures [33], [34]. The model includes a full-package and air cooling solution for the face-to-back integrated stack. A range of thermal conductivities were explored for the TSV structures including copper, tungsten, and composite cases (with spanning a range of best case copper to pessimistic cases dominated by insulator conductivities)(Table III in Section VI). TSV pitch range of 2 − 5 $\mu$m was investigated. Individual vias are modeled for the TSV farms including the electrical passivation. Table I summarizes the material conductivities in the stack [2]. We also investigated a range of TSV alternatives with: 1) varying conductivity ranges; 2) with and without electrical insulation; and 3) via pitch and size alternatives.

### B. Thermal Modeling of the TSV-Farm Blocks

The TSV farms were modeled such that they have independent lateral and vertical thermal conductivity components, which improves the accuracy of representation over the traditional way of only focusing on vertical conductivity (in all prior art). The TSV farms still act as goods conductor between

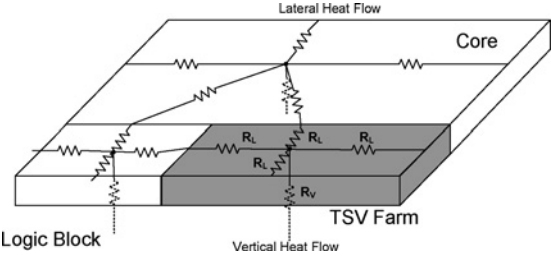

Fig. 4. Block mode thermal RC modeling for TSV-farm structures. Each block is modeled with a set of thermal resistances.

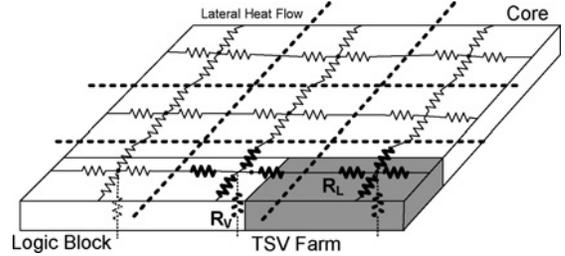

Fig. 5. Grid level thermal RC modeling for TSV-farm structures. The whole die is equally divided into grid cells. For each cell, $R_L$ and $R_V$ stands for the composite thermal resistance values in the lateral and the vertical direction, respectively.

The RC model consists of a vertical and a lateral convection model. Fig. 4 shows the block level RC modeling, where a TSV-farm region is modeled as a block with its own thermal resistances. The vertical thermal resistance ($R_V$) captures heat flow from one layer to the next, moving from the source die through the package. The lateral thermal resistance ($R_L$) captures heat diffusion between adjacent blocks within a layer, and from the edge of one layer into the periphery of the next area. The thermal resistance $R$ is proportional to the thickness of the material ($h$) and inversely proportional to the cross sectional area ($A$) across which the heat is being transferred

$$R = \frac{h}{k \cdot A} \qquad (1)$$

where $k$ is the thermal conductivity of the material per unit volume. For TSV-farm structures, note that the lateral and vertical thermal conductivities are different, according to the discussion in previous sections. The lateral and vertical resistances for the TSV-farm structures are then calculated as

$$R_L = \frac{h_L}{k_{\text{farm}} \cdot A_L}, \quad R_V = \frac{h_V}{k_{\text{metal}} \cdot A_V}. \qquad (2)$$

Block level RC modeling works in a coarse granularity. To improve accuracy of thermal analysis, a grid-like model is used, which extends the original block-level model and enables thermal analysis at arbitrary granularity. Fig. 5 shows the modeling on a 3 × 3 grid. Each grid cell has its own thermal resistances and power dissipation values. While the TSV-farm structures are taken into account in the grid model, a grid cell may consist of both silicon materials and TSV-farm

structures. In this case, the thermal resistances ($R_L$ and $R_V$) are calculated as

$$R_{L,\text{cell}} = \frac{h_L}{k_{\text{farm}} \cdot \eta_{\text{farm}} \cdot A_L} + \frac{h_L}{k_{Si} \cdot \eta_{Si} \cdot A_L} \quad (3)$$

$$R_{V,\text{cell}} = \frac{h_V}{k_{\text{metal}} \cdot \eta_{\text{farm}} \cdot A_V} + \frac{h_V}{k_{Si} \cdot \eta_{Si} \cdot A_V} \quad (4)$$

where $k_{\text{farm}}$ is obtained from the characterization presented in previous sub-section. $A_L$ and $A_V$ are the cross-sectional areas in the lateral and vertical directions, respectively. $\eta_{\text{farm}}$ and $\eta_{Si}$ are the ratios of areas each material occupies in the grid cell.

The grid size is predetermined and based on the technology specifications (specifically to capture the minimum feature size for placement accurately—in this case vias sizes /pitch). With the proposed thermal RC modeling, HotSpot [39] is called to calculate the temperature numbers of each grid in the stack. HotSpot is embedded in our framework as an in-place thermal estimator and the accuracy has been validated against commercial thermal analysis tools.

## V. THERMAL-BLOCKAGE-AWARE TSV-FARM OPTIMIZATION

In this section, we present the design methodology to optimize the thermal characteristics of TSV-farm (3-D buses).

### A. TSV-Farm Optimization Objective

The heat conduction $H_{AB}$ between two units A and B on the same layer can be expressed as

$$H_{AB} = \Delta Q / \Delta t = k \cdot A \cdot \Delta T / x \quad (5)$$

where $k$ is thermal conductivity of the material, $A$ is the cross-sectional area, $x$ is the distance between blocks A and B, and $\Delta T$ is the temperature difference between the units. In this context, the $\Delta Q / \Delta t$ is the rate of heat flow between A and B. When the TSV-farm structures are taken into account, a new metric, which is defined as the heat conduction at a unit temperature difference, is introduced to quantify the heat conduction efficiency with different thermal conductivities

$$f_{H_{AB}} = \Delta Q / (\Delta T \cdot \Delta t) = k \cdot A / x. \quad (6)$$

The heat conduction efficiency $f_{H_{AB}}$ captures the impact of TSV-farm structures on thermal via thermal conductivity k and the geometrical parameters A and x. As the cross-sectional area A can be reduced significantly (from 700−800μm silicon to 10 − 20μm silicon), the resulting lateral heat conduction is proportionally affected in thinned silicon. For a given distance x between blocks, our proposed technique improves the heat conduction efficiency $f_{H_{AB}}$ by enabling the heat dissipation along the path with higher thermal conductivity and larger contact area. In particular, the algorithm maximizes the heat conduction efficiency on the maximum thermal gradients, where the temperature distance between the source and drain are highest. By arranging the geometries (in terms of aspect ratio) of TSV-farm structures, along with replacement, the algorithm improves the thermal profile of chips in the challenging lateral direction.

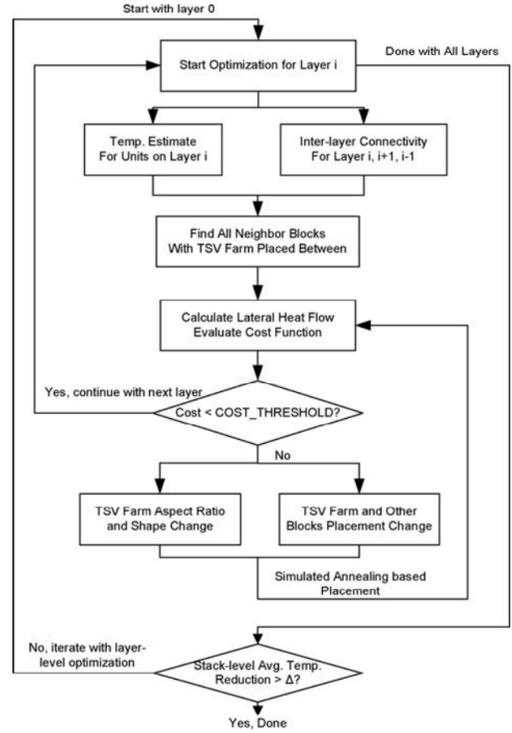

Fig. 6. Details of the temperature-aware vertical interconnect optimization technique.

### B. TSV-Farm Optimization Flow

Optimization at the floorplanning level could bring in significant improvements [42], [43]. The proposed flow optimizes the performance and area driven solution by incorporating temperature awareness during floorplanning. Details of the temperature-aware bus placement flow are illustrated in Fig. 6. In the first stage, the initial thermal profiles of the individual units are obtained from design time thermal analysis, along with the vertical interconnectivity among layers in the stack,

The heat conduction efficiency $f_{H_{AB}}$ captures the impact of TSV-farm structures on thermal via thermal conductivity $k$ and the geometrical parameters $A$ and $x$. As the cross-sectional area $A$ can be reduced significantly (from 700−800$\mu$m silicon to 10 – 20$\mu$m silicon), the resulting lateral heat conduction is proportionally affected in thinned silicon. For a given distance $x$ between blocks, our proposed technique improves the heat conduction efficiency $f_{H_{AB}}$ by enabling the heat dissipation along the path with higher thermal conductivity and larger contact area. In particular, the algorithm maximizes the heat conduction efficiency on the maximum thermal gradients, where the temperature distance between the source and drain are highest. By arranging the geometries (in terms of aspect ratio) of TSV-farm structures, along with replacement, the algorithm improves the thermal profile of chips in the challenging lateral direction.

The algorithm features two loops of iterations: one at stack level and one at layer level. In the inner loop, the algorithm follows a bottom-up layer-by-layer order. In each layer $i$, the algorithm estimates the initial thermal profile, and finds out the TSV-farm structures that start from layer $i$.

The TSV-farm structures are treated as soft blocks in the view of floorplanning. A set of candidate values for the aspect ratio change are pre-defined, with which thermal characterizations are done to extract the conductivities. When the aspect ratio needs to be changed for optimization, a value from the pre-set candidates is chosen so that the expected thermal resistance in the flow direction is minimized.

A simulated-annealing based floorplanning is then performed. Several optimization moves can be applied to explore the design space, including: 1) Replace longer/higher aspect ratio bus geometries with wider/square alternatives; and 2) moving the bus structures around the planar to improve the thermal conduction.

```
SAPLACEMENT(S)
    ▷ Initialization
 1  Get an initial floorplan S_Best = S;
 2  Get an initial temperature T > 0;
    ▷ Simulated-annealing based placement
 3  while T > T_THRESHOLD
 4      do for i ← 1 to MAX_MOVES
 5          do S' ← GENMOVE(S)
 6              ΔC = cost(S') - cost(S);
 7              if ΔC ≤ 0
 8                  do S ← S'; ▷ Down-hill move
 9              else
10                  S ← S' with the prob. of e^{-ΔC/T};
11              if cost(S_Best) > cost(S)
12                  do S_Best ← S
13          T = rT;
14  Return S_Best;

GENMOVE(S)
 1  Randomly pick up a TSV-farm structure from S;
 2  Generate a random number i between 0 and 1;
 3  if i < 1/2
 4      do Change the aspect ratio;
 5          Recompute the width/height of the TSV-farm;
 6  else
 7      Change the position of TSV-farm in the floorplan;
 8  Recompile the floorplan S;
 9  Return S;
```

Fig. 7. Outline of the simulated-annealing based placement algorithm.

The floorplanner picks one of the moves for each inner iteration to ensure that the change is localized and the optimization can converge. A cost function incorporating multiple evaluation metrics is used to guide the floorplanning. After the optimization in layer $i$ is finished, the algorithm enters a new iteration with layer $i + 1$ until all layers are optimized. The outer loop then checks whether the average temperature of the whole stack is reduced by the layer level optimization. The algorithm terminates after a given number of iterations and the floorplannings leading to the maximum overall reduction is recorded.

C. *Simulated-Annealing Based Placement of TSV-Farm Structures*

A simulated-annealing based floorplanner is employed during the optimization to facilitate the changes of aspect ratio and placement of TSV-farm structures. The outline of the algorithm is shown in Fig. 7, which starts with a generic simulated annealing framework (Line 3–13 of the *SAPlacement* routine), and iteratively changes the aspect ratio of TSV-farm structures (Line 4–5 of the *GenMove* routine) as well as moves the structures around (Line 7 of the *GenMove* routine), to obtain a better thermal distribution across the chip.

In Fig. 7, $S$, $S'$, and $S_{Best}$ denote the current floorplan, the alternative floorplan, and the best floorplan obtained, respectively. $T$ is the annealing temperature that is featured in the simulated-annealing optimization. T _ THRESHOLD and $r$ are parameters to control the simulated-annealing optimization. T THRESHOLD is the temperature threshold. The annealing schedule will terminate if temperature goes below the threshold. r is the cooling down factor. By tuning both of the parameters, the trade-off between total run time and quality of results of the optimization can be balanced.

The cost function used during the placement algorithm (Line 6 and 9 in the SAPlacement routine in Fig. 7) is defined as follows:

$$F_{cost} = \alpha \cdot A + \beta \cdot f_H + \gamma \cdot R + \delta \cdot W \quad (7)$$

where

$$f_H = \sum_{A, B \text{ are adjacent}} f_{H_{AB}} \quad (8)$$

$$R = |Ratio - Ratio_{preset}| \quad (9)$$

where $\alpha$, $\beta$, $\gamma$, and $\delta$ are weighting factors. $A$ is the final floorplan area, $f_H$ is the total heat gradient efficiency involving TSV-farm structures, $R$ is to regulate the overall aspect ratio of the floorplan, and $W$ is the total wirelength.

First, in order to bound the floorplan overhead on silicon area, $A$ and $R$ are included in the cost function. The replacement of TSV-farm structures will also lead to a redistribution of interconnect wires in the metal layers. Therefore, the floorplanner uses a simple wire length model based on manhattan distance, to estimate the wire length overhead.

The heat gradient efficiency function is the major metric to guide the floorplanning. As higher efficiency is desired, the weighting factor for the cost contribution of $f_H$ should be negative. For the weighting factors of $A$ and $W$, the values are chosen to be considerably high to impose high penalties on area or wire length overheads. This ensures the TSV-farm optimization with improvements on thermal and negligible overhead on chip area and total wire length.

The cost function does not cover timing and power estimations, with the observation that the proposed technique makes thermally-aware changes in the TSV farm placement, such that with minimum changes in the placement and aspect ratio temperature improvement is achieved in all of the experimental cases. We observed this trend for all the cases considered. While it is true that there might be special circumstances at macro or unit levels under which timing or power degradation is non-negligible, those are out of the scope of the study (for which the focus is at core-level). It's also worth mentioning that tool considers a range of aspect ratio and placement changes to pick the minimum wirelength and temperature option among the selections within specified ranges. It does not make major placement moves that would alter the timing or power dissipation.

## V. EXPERIMENTAL RESULTS

In this section, we explore the effectiveness of the vertical bus structure (or TSV-farm) optimization technique in 3-D designs with either unit-level or core-level partitioning. We also investigate the impact of varying device layer counts and thermal conductivities on the effectiveness of the optimization. Table III lists the technology parameters used in the experiments. The average values in the ranges are used as default while full ranges are explored for experiments. For modeling of power dissipation, unit-level power values are used for each layer, in addition to factoring in the temperature of the unit (thermal induced leakage effects are calculated accordingly) to get the total power dissipation. The thermal evaluation is performed in the modeling infrastructure as discussed in Section IV.

TABLE III
TECHNOLOGY PARAMETERS USED IN THE EXPERIMENTS

| Parameter | Value |
|---|---|
| TSV farm thermal conductivity | $0.5 - 5 W/(mK)$ (Range Explored) |
| TSV size/pitch | $2 - 5\ \mu m@2x$ [44], [45] |
| Avg. TSVs per core | $10^4$ |
| Average core area | $10\ mm^2$ |
| Power density | $1 - 5\ W/mm^2$ |
| Silicon thickness | $10 - 100\ \mu m$ thin Si |
| Ambient temperature | $25^oC$ |

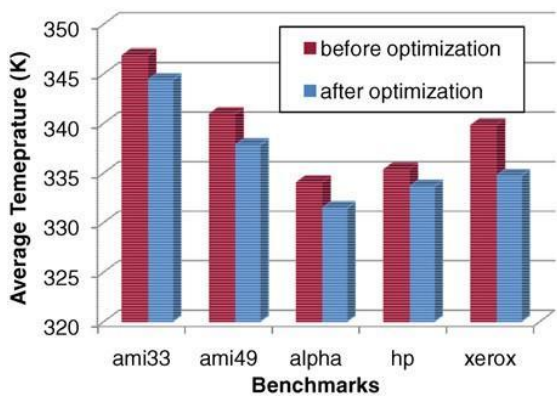

Fig. 8. Average temperature reduction achieved with the thermal aware vertical bus optimization flow. The *x*-axis shows the benchmark set and the *y*-axis shows the average temperature of the chip.

### A. Thermal Optimization Results for Unit-Level Partitioning

To quantify the temperature reduction brought by the TSVfarm optimization, we conduct experiments on a set of MCNC benchmarks. The benchmark circuits are partitioned at the unit-level, using a similar approach in [29], to minimize total interconnect wire length as well as the peak temperature. The interconnect information is extracted from the initial floorplan and the TSV-farm blocks are created accordingly.

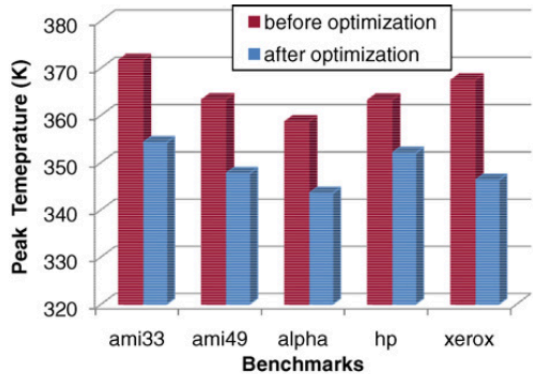

Fig. 9. Peak temperature reduction achieved with the thermal aware vertical bus optimization flow. The *x*-axis shows the benchmark set and the *y*-axis shows the peak temperature of the chip.

We then apply the TSV-farm optimization flow on the benchmark circuits, and the results are reported in Table II. Columns 2–3 and 7–8 show the wire length and the chip area before and after the TSV-farm optimization, while columns 4–6 and 9–12 compare three temperature values: the average temperature all over the chip (avgT), the peak temperature (peakT), and the average temperature of the hottest block in the chip, respectively. Figs. 8 and 9 depict the reduction of average and peak temperatures in detail. On average a 16 K reduction on chip peak temperature and 3 K reduction on chip average temperature are achieved with negligible wire length overhead, and the chip area is slightly reduced due to the reshape and replacement of TSV-farm structures. We can see that although the input circuits are already optimized for peak temperature minimization, a significant improvement of thermal profile can still be obtained with the proposed optimization flow, given that the lateral thermal conduction of TSV-farm structures is taken into account.

### B. Core-Level Case Study 1: Multi-Core Stacking

After examining the cases for unit-level partitioning, we perform a set of case studies on 3-D bus based architectures at core-level partitioning. Core-level integration is operating at a coarser granularity with much larger silicon area and more intensive use of TSV-farm structures as vertical bus connections.

Fig. 10 shows the floorplan of a stacked multicore architecture used for the first case study. We consider a 2-

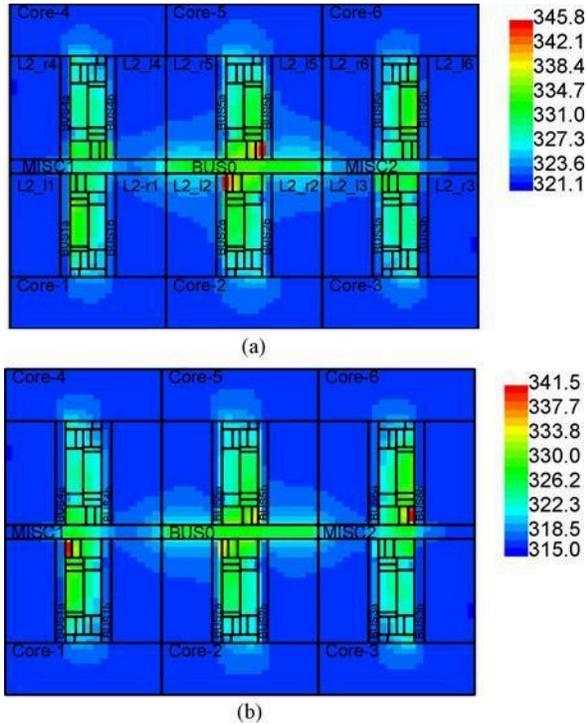

Fig. 10. Thermal profile of the multicore architecture before the thermal aware bus optimization. (a) Thermal map of the bottom layer. (b) Thermal map of the top layer. Note that the thermal maps are in different scales.

layer multicore stack, where each core has its private L2 cache, with a set of vertical bus architectures going (from the I/O pads) through the bottom layer and connecting to the layer immediately on top. Each layer has six SPARClike processor cores with surrounding TSV-farm structures, one shared bus structure `BUS0`, as well as several blocks for peripheral circuits, as shown in Fig. 10. The floorplan of the shown layer is repeated in the top layer.

We first show in Fig. 10 the temperature maps of the two layers. The figures were generated using the Hotspot thermal analysis tool, which is a well-recognized thermal simulation tool that supports multilayer simulation of 3-D ICs. The figures represent 3-D simulations of multiple layers. In the bottom layer as shown in Fig. 10(a), local hotspots are formed in the execution units of `CORE2` and `CORE5`. Such a steep temperature gradient across the TSV-farm structures is uncommon in conventional thermal profiles when the lateral heat blockage of TSV-farm is not taken into account. However, local hotspots are not formed near the TSV bus structures in the top layer [Fig. 10(b)] because TSV farms in this layer are connected to the metals and the blockage effect is eliminated.

The thermal-aware bus structure optimization flow is then performed and the optimization results are shown in Fig. 11(a) and (b), where the TSV-farm structures are reshaped and repositioned, leading to a slightly different floorplan. From the figure, we can see that the local hotspots are migrated with much lower peak temperatures. Note that the average temperatures of the two layers are changing in different directions, as shown in Table IV. The optimization algorithm is able to trade off the improvement in one layer to degradation

TABLE IV

CHANGES IN AVERAGE TEMPERATURES OF EACH LAYER

| Avg. Temp. (K) | Bottom Layer | Top Layer | Whole Stack |
|---|---|---|---|
| Before Opt. | 323.58 | 319.87 | 321.71 |
| After Opt. | 319.06 | 320.35 | 319.70 |

TABLE V

COMPARISON OF TEMPERATURES BETWEEN CORE-LEVEL HOTTEST BLOCKS BEFORE AND AFTER OPTIMIZATION

| T(K) | Peak | C1 | C2 | C3 | C4 | C5 | C6 |
|---|---|---|---|---|---|---|---|
| Before | 346.8 | 335.8 | 339.2 | 335.9 | 335.9 | 339.2 | 335.8 |
| After | 334.0 | 331.8 | 332.0 | 331.8 | 331.8 | 332.0 | 331.8 |
| Reduction | 12.8 | 4.0 | 7.2 | 4.1 | 4.1 | 7.2 | 4.0 |

TABLE VI

COMPARISON OF TEMPERATURES BETWEEN CORRESPONDING BLOCKS BEFORE AND AFTER OPTIMIZATION

| T(K) | IntExec1 | IntQ1 | IntReg1 | IntExec4 | IntQ4 | IntReg4 |
|---|---|---|---|---|---|---|
| Before | 331.6 | 332.4 | 335.8 | 333.3 | 335.0 | 337.2 |
| After | 331.0 | 328.9 | 329.8 | 330.2 | 330.7 | 331.8 |
| Reduction | 0.6 | 3.5 | 5.0 | 3.1 | 4.3 | 5.4 |

in the other layer, and to achieve a reduction in the average temperature of the whole stack.

Table V compares the temperature of local hotspots on each core of the bottom layer before and after the optimization. Column 2 shows the peak temperature of the chip, while columns 3–8 show the temperatures of the hottest blocks in `CORE1–CORE6`, denoted by `C1–C6`, respectively. As shown in the table, the peak temperature is reduced by 12 K due to the removal of lateral heat blockage of TSV-farm structures. The average temperatures of the hottest blocks on each core are reduced to different extents, while `CORE2` and `CORE5` benefit most from the TSV-farm optimization, with their hottest blocks located in the center of the chip and the thermal blockage of both local BUS structures and the shared `BUS0` now eliminated.

To further examine the thermal blockage effect of TSV-farm structures with more detailed information, we zoom in the views in Figs. 10 and 11, and focus on the crossing region between `CORE1` and `CORE4`. The detailed thermal profile on block-level of the region is shown in Fig. 12, and the temperature numbers are listed in Table VI. As shown in Fig. 12(a), the TSV-farm structures `BUS1a`, `BUS1b`, `BUS4a`, `BUS4b` are surrounding the logic blocks, and `IntReg1` and `IntReg4` become local hotspots (according to the temperature numbers shown in Table VI) due to the thermal blockage of the TSV-farms. After the TSV-farm placement optimization, the TSV-farms are replaced with BUS0a, leading to a redistribution of the thermal profile as shown in

Fig. 12(b). The average temperature across all the blocks is reduced according to Table VI, with the thermal blockages of TSV-farms on the x- axis direction eliminated and heat dissipating through the L2 blocks. However, as BUS0a blocks part of the heat conduction in the y-axis direction, thermal dissipation of blocks around BUS0a are affected. As a result, IntExec1 becomes a new

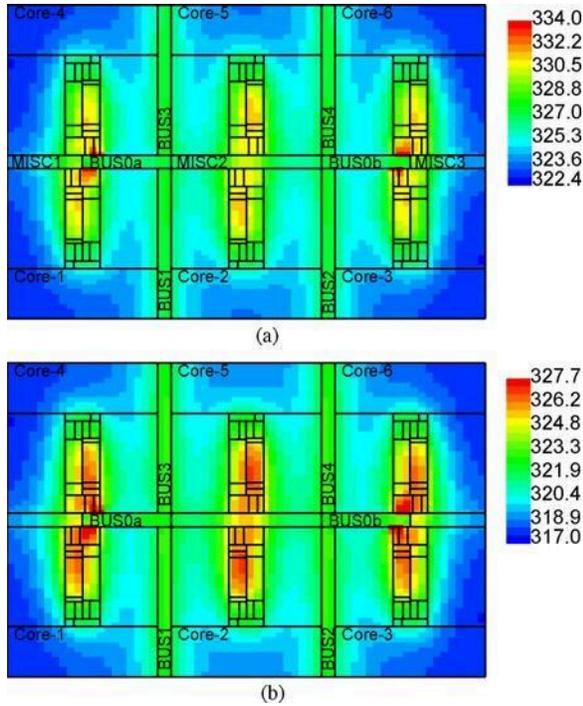

Fig. 11. Thermal profile of the multicore architecture after the thermal-aware bus optimization. (a) Thermal map of the bottom layer. (b) Thermal map of the top layer. Note that the thermal maps are in different scales.

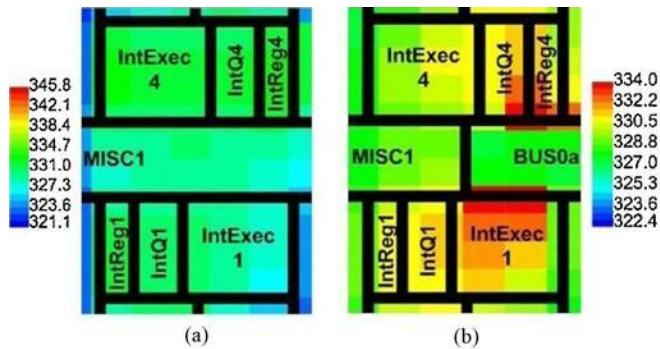

*Please note the color scale differences - option (b) is cooler than option (a)

Fig. 12. Zoomed-in view of the multicore architecture showing the blockage effect of TSV-farm structure. (a) Before thermal-aware bus optimization. (b) After the thermal-aware bus optimization.

local hotspot instead of IntReg1. This table highlights the importance and necessity of the proposed design planning approach in 3-D. Thermal aware floorplanning techniques may not be effective without careful placement/planning with respect to the surrounding blockages.

*C. Core-Level Case Study 2: MultiCore + Memory Stacking*

In the second case study we look at a thermally aggressive stacking case with 1–4 layers of cache stacked on a core layer. As memory blocks may come from different vendors, we assume that the on-chip cache blocks have at least the default granularity that our tool sets and can be integrated in the main design without external vendor constraints.

In previous work on cache and memory stacking, multiple layers stacked on the processor obstruct the vertical heat local hotspot instead of IntReg1. This table highlights the importance and necessity of the proposed design planning approach in 3-D. Thermal aware floorplanning techniques may not be effective without careful placement/planning with respect to the surrounding blockages.

*C. Core-Level Case Study 2: MultiCore + Memory Stacking*

In the second case study we look at a thermally aggressive stacking case with 1–4 layers of cache stacked on a core layer. As memory blocks may come from different vendors, we assume that the on-chip cache blocks have at least the default granularity that our tool sets and can be integrated in the main design without external vendor constraints.

In previous work on cache and memory stacking, multiple layers stacked on the processor obstruct the vertical heat conduction and exacerbate the thermal problem in 3-D stacks. In consequence the heat dissipation of the processor layer rely more on the lateral heat conduction and the impact of TSVfarm blockage might be more significant.

In this case study, the thinned processor layer is assumed to be face-down, close to the package. Such configurations are desirable for efficient power delivery and IO connections. As a result the 3-D interconnect structures connecting to the memory layers obstruct the heating.

Compared with the case of two-layer core-stack architecture, the peak and average temperatures increase significantly, due to the vertical thermal obstruction of the stacked cache layers. The thermal-aware TSV bus optimization flow is then performed on the whole stack. The thermal profiles before and after optimization for the four-layer stacking case are shown in Figs. 13(a) and (b), respectively. Fig. 13(c) depicts the thermal reduction brought by the proposed optimization, where the thermal gradient is reduced as the TSV-farm structures are moved around. The peak temperature of the core layer is reduced by 10.7 K, and the average temperatures of the core layer and the whole stack are reduced by 5.3 K and 2.8 K, respectively.

Fig. 13. Thermal aware TSV bus optimization for the core-memory architecture with four layers of stacked memory. (a) Thermal map of the core layer before the optimization. (b) Thermal map of the core layer after the optimization. and (c) shows the temperature differences between (a) and (b). Note that the thermal maps are in different scales. (a) Before bus optimization.
(b) After bus optimization. (c) Temperature reduction achieved.

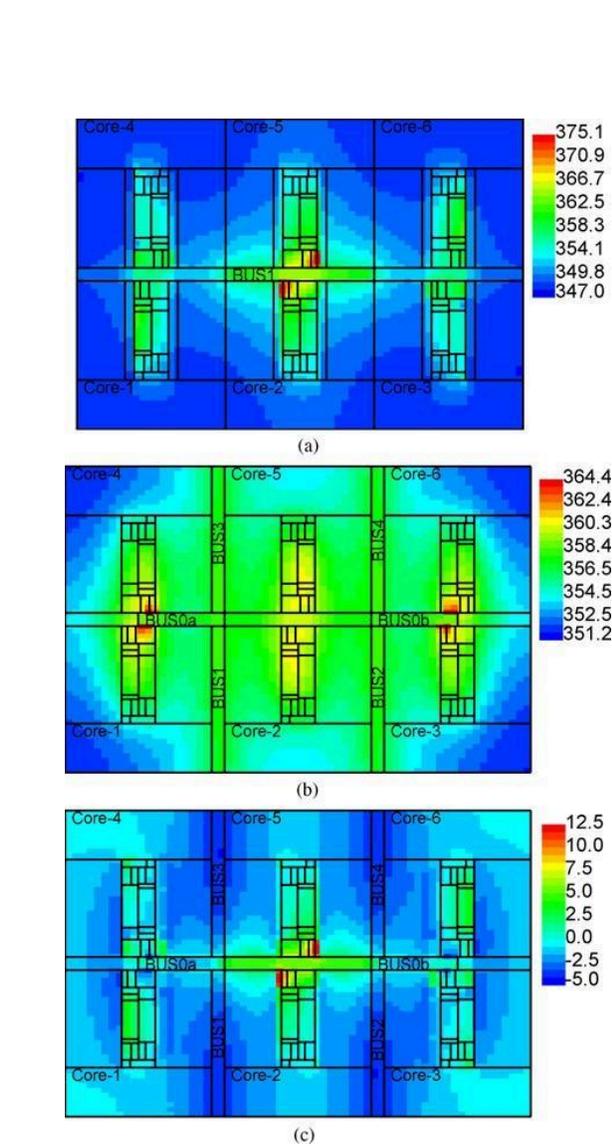

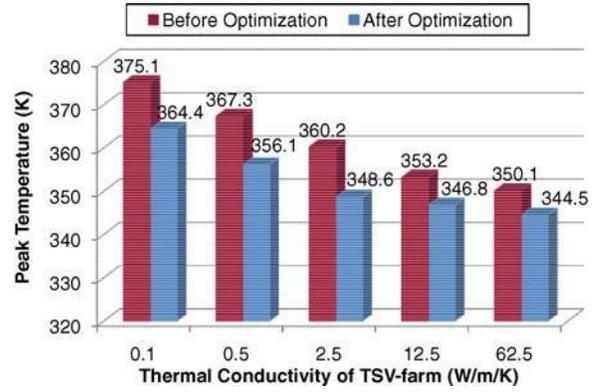

Fig. 14. Peak temperature reduction with different numbers of stacked memory layers. The $x$-axis shows the number of memory layers stacked on the core layer. The $y$-axis with the data labels shows the peak temperature of the core layer for each case.

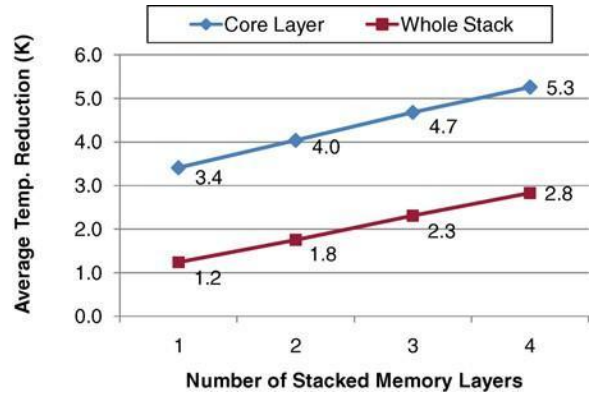

Fig. 15. Average temperature reduction with different numbers of stacked memory layers. The $x$-axis shows the number of memory layers stacked on the core layer. The $y$-axis with the data labels shows the average temperature values.

1) *Impact of Layer Counts in the Stack:* To further explore the impact of layers counts on the effectiveness of our proposed optimization technique, we vary the number of stacked memory layers in the stack and compare the results in Figs. 14 and 15. In Fig. 14, the $x$-axis shows the number of memory layers stacked on the core layer. Note that up to four layers of stacked memory are explored, corresponding to the typical maximum layer counts due to the heat dissipation challenge. The $y$-axis with the data labels shows the peak temperature of the core layer for each case. The red (dark) bar and blue (light) bar show the values before and after the proposed optimization, respectively.

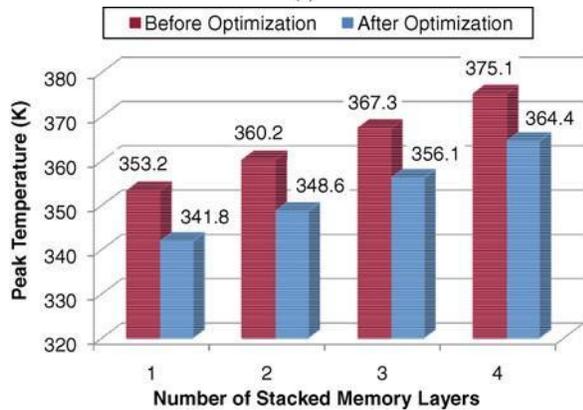

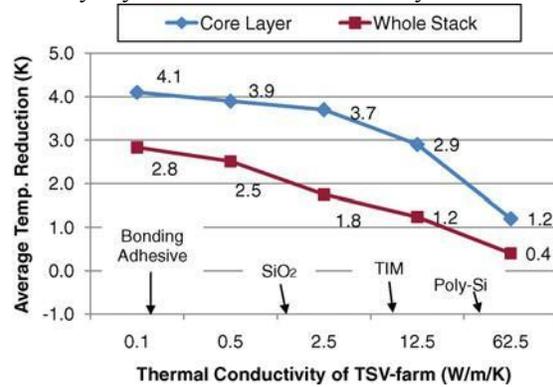

As the number of stacked memory layers increases, the peak temperature of the core layer steps up, and the proposed optimization is able to bring in a steady reduction about 11 K. For the average temperature reduction in Fig. 15, the blue

Fig. 16. Peak temperature reduction with different thermal resistances of the TSV-farm structure. The $x$-axis shows different thermal resistance values of the TSV-farm structure. The $y$-axis with the data labels shows the peak temperature of the core layer for each case.

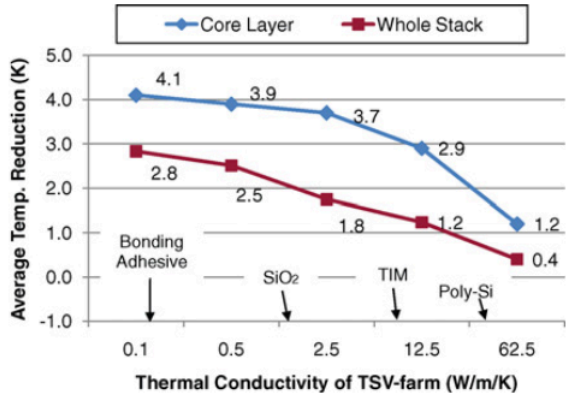

Fig. 17. Average temperature reduction with different thermal resistances of the TSV-farm structure. The $x$-axis shows different thermal resistance values of the TSV-farm structure. The $y$-axis with the data labels shows the average temperature values.

(light) curve and the red (dark) curve show the values for the whole stack and for the core layer, respectively. Fig. 15 shows a nearly linear dependency of the temperature reduction on the layer counts, which means the reduction is higher on 3-D stacks with more memory layers.

*2) Impact of Thermal Conductivity Values of the TSVFarm Structure:* To explore the impact of thermal conductivity modeling on the effectiveness of our proposed optimization technique, we vary the thermal conductivity values of the TSVfarm structure and compare the results in Figs. 16 and 17. In Fig. 16, the $x$-axis shows different thermal conductivity values of the TSV-farm structure. Note that the $x$-axis is log-scale and the range explored covers most of the materials used in semiconductor manufacture. The $y$-axis with the data labels shows the peak temperature of the core layer for each case. The red (dark) bar and blue (light) bars show the values before and after the proposed optimization, respectively.

With lower thermal conductivity of the TSV-farm structure, the lateral thermal blockage effect is more prominent, therefore, the peak temperature of the core layer is much higher as shown in Fig. 16, and the reduction brought by the proposed optimization is more significant. For the average temperature reduction in Fig. 17, the blue (light) curve and the red (dark) curve show the values for the whole stack and for the core layer, respectively. Note that the $x$-axis is logscale and the labels show the thermal conductivity values of the materials used in semiconductor manufacture, where TIM stands for thermal interface material. Fig. 17 demonstrates that as the thermal conductivity value approaches those of poly-silicon and silicon, the average temperature reduction diminishes because of the declining of the lateral thermal blockage effect.

## VI. CONCLUSIONS

3-D integration has been shown to provide significant advantages in terms of interlayer bandwidth and reduced latency. Yet the dense via farm structures of the 3-D bus that are used for signal transmission can exacerbate the existing thermal challenges in 3-D IC design. Our experimental results showed that TSV farms can cause different thermal effects on different layers due to the unequal x, y, z thermal conductivities. This can exhibit itself as thermal improvement in the vertical heat flow in addition to lateral heat blockage effects in passthrough layers. We proposed a temperature-aware TSV-farm (3-D bus) architecture flow, which minimizes the impact of vertical interconnect, by improving the heat conduction in the lateral direction through via-farm adjustments. Our experimental results indicated that the temperature-aware 3-D bus flow improves the thermal profile of the 3-D IC designs.

TABLE II

TSV-FARM OPTIMIZATION ON A SET OF MCNC BENCHMARKS

| Circuit | Before TSV-Farm Optimization | | | | | After TSV-Farm Optimization | | | | | |
| --- | --- | --- | --- | --- | --- | --- | --- | --- | --- | --- | --- |
| | wire length ($\mu$m) | area ($mm^2$) | avgT (K) | peakT (K) | hottest blk(K) | wire length ($\mu$m) | area ($mm^2$) | avgT (K) | peakT (K) | hottest blk (K) | run time (s) |
| ami33 | 133323 | 4.84 | 346.83 | 371.65 | 350.51 | 133539 | 4.8 | 344.33 | 354.17 | 345.17 | 85 |
| ami49 | 2540451 | 149.72 | 340.91 | 363.30 | 343.19 | 2540475 | 145 | 337.77 | 347.63 | 339.25 | 1304 |
| alpha | 1019016 | 117.72 | 334.02 | 358.63 | 338.03 | 1019490 | 115.12 | 331.41 | 343.44 | 334.40 | 856 |
| hp | 399606 | 35.8 | 335.32 | 363.22 | 341.46 | 400527 | 33.92 | 333.59 | 351.95 | 341.36 | 155 |
| xerox | 1628778 | 78.76 | 339.79 | 367.41 | 345.59 | 1631565 | 76.52 | 334.67 | 346.18 | 337.35 | 640 |
| Average | 100% | 100% | 339.37 | 364.84 | 343.76 | **101%** | **98%** | 336.35 | 348.67 | 339.50 | - |